\documentclass[RNAAS]{aastex631}

\defcitealias{Noirot2022}{N22}

\shorttitle{Noirot et al.}
\shortauthors{Noirot et al.}
\graphicspath{{./}{./}}

\begin{document}

\title{H$\alpha$ AND CONTINUUM SIZES WITH THE {\it HST}/WFC3 G141 GRISM: OUTSIDE-IN QUENCHING FOR $z=1.0-1.4$ FAST QUENCHERS?}

\author{Ga\"el Noirot}
\affiliation{
Department of Astronomy \& Physics, Saint Mary’s University, 923 Robie Street, Halifax, NS B3H3C3, Canada}

\author{Marcin Sawicki}
\affiliation{
Department of Astronomy \& Physics, Saint Mary’s University, 923 Robie Street, Halifax, NS B3H3C3, Canada}



\begin{abstract}

We investigate the evolution of the physical extent of star formation of $M_{\star}>10^9~M_{\odot}$ rapidly-quenching galaxies at $z=1.0-1.4$. We measure the galaxy H$\alpha$ and stellar continuum sizes from their {\it HST}/WFC3 G141 grism spectroscopy and connect the galaxy sizes to time on their evolutionary delayed-$\tau$ tracks determined in \citet{Noirot2022}. Most galaxies (10/13) have non-evolving H$\alpha$-to-continuum size-ratios consistent with unity within the measurement uncertainties, suggesting an homogeneous decline of star-formation in these galaxies despite a rapid shut-down of their star formation.  On the other hand, a handful (3/13) show statistically smaller H$\alpha$ sizes compared to the stellar continuum as they age and approach the blue-cloud/red-sequence transition region. This suggests an outside-in shut-down of the star-formation (potentially driven by environmental mechanisms) in these rapidly evolving galaxies as they move from the blue cloud towards the red sequence.

\end{abstract}



\section{Introduction} \label{sec:intro}
	
	The H$\alpha$ emission line ($\lambda_{\rm rest} = 6563$~\AA) is a gold-standard tracer of recent star formation in galaxies (e.g., \citealp{Kennicutt98}). The spatial distribution of the H$\alpha$ line, when compared to the (older) stellar continuum, can tell us where galaxies form their stars over time and under different conditions. 
	Several studies have now demonstrated the feasibility of this approach using {\it HST}/WFC3 grism spectroscopy and associated broad-band imaging 
	(e.g., \citealp{Nelson2012,Nelson2016,Matharu2021,Matharu2022}). Using stacking analyses, these studies showed notably that $z\sim1$ field galaxies experience an inside-out growth of their stellar component followed by the inside-out quenching of the formed stellar disk, in line with the size-mass relation of star-forming field galaxies. Using a similar stacking approach, they also showed that recently and rapidly quenched galaxies in 
	a sample of $z\sim1$ clusters 
	experience rapid outside-in disk truncation, 
	consistent with environmental processes such as ram-pressure stripping. However, the link between galaxy size evolution and individual galaxy star-formation histories (SFHs)/evolutionary sequences remains to be investigated.
	 	
	 	
	In \citet[hereafter \citetalias{Noirot2022}]{Noirot2022} we perform the spectral energy distribution fitting of deep {\it HST}/WFC3 G102 \& G141 grism spectroscopy to study the evolution of $z\sim1.0-1.8$ galaxies from the photometric blue cloud, across the green valley, and into the red sequence. That work identifies three distinct galaxy populations differentiated by their delayed-$\tau$ (i.e., delayed exponential) timescales: fast ($\tau<0.5$~Gyrs), intermediate ($0.5<\tau<1.5$~Gyrs), and slow ($\tau>1.5$~Gyrs). At these redshifts, only the fast-quenching galaxies have had enough time to go through the full quenching sequence of blue cloud $\rightarrow$ green valley $\rightarrow$ red sequence. Focusing on this $\tau<0.5$~Gyr population allows us to explore the full evolutionary life-cycle of quenching galaxies at high redshifts. In \citetalias{Noirot2022}, we examine how fast these rapidly-quenching galaxies cross the green valley ($\sim$1~Gyr) and find that their numbers and the rate at which they migrate across the green valley are sufficient to fully account for the build-up of the quiescent galaxy population over time. Here, we turn to size evolution, and have a first look at how the physical extent of star formation changes in these fast-quenching galaxies as they shut down their star formation on the way from the blue cloud towards the red sequence.

\section{Spatial Profiles} \label{sec:profiles}

	Our initial galaxy sample consists of the $z=1.0-1.8$ 
	fast-$\tau$ population from \citetalias{Noirot2022} (65 sources, including 37 blue-cloud, 12 green-valley, and 16 red-sequence galaxies).  From this sample, we select sources with observed H$\alpha$ emission 
	in the G141 grism data. 
	Although this mostly restricts our sample to blue-cloud galaxies, the galaxies are seen at various times past their peak of star formation, which constitutes different stages of their evolution towards quiescence (Figure~\ref{fig:grismdata}(b)). From the drizzled 2D G141 data, we extract the line and continuum spatial profiles 
	within boxes of $\pm1000$ and $\pm5000$ km/s rest-frame, 
	respectively. We center the line extraction boxes on 
	H$\alpha$ 
	and 
	the continuum extraction boxes on two regions on 
	each sides of the 
	lines (see Figure~\ref{fig:grismdata}(a)). 
	We then average-stack the 2D regions along the wavelength direction to create 1D line and continuum spatial profiles, 
	and subtract the 1D continua from the lines to obtain the H$\alpha$ spatial profiles (see the blue --H$\alpha$-- and red --continuum-- histograms in Figure~\ref{fig:grismdata}(a)).
	
	We fit the H$\alpha$ and continuum profile of each source with single Gaussians, 
	and measure the profiles' full width half maximum (FWHM) and the H$\alpha$-to-continuum FWHM ratios ($\rm FWHM_{H\alpha}/FWHM_{Conti}$) from the fits. 
	To derive our measurement uncertainties, each profile is perturbed at the pixel level following the pixel flux uncertainties and fitted 500 times. 
	After applying a FWHM threshold of $>0.25\arcsec$, our final sample consists of 13 galaxies within $z=1.0-1.4$, $\log_{10}(M_{\star}/M_{\odot}) = 9 - 10.8$, and delayed-$\tau$ characteristic timescales of $\tau = 0.1-0.35$~Gyrs.

\section{Size Evolution} \label{sec:sizeevol}

\begin{figure}[!ht]
\includegraphics[width=1.035\textwidth]{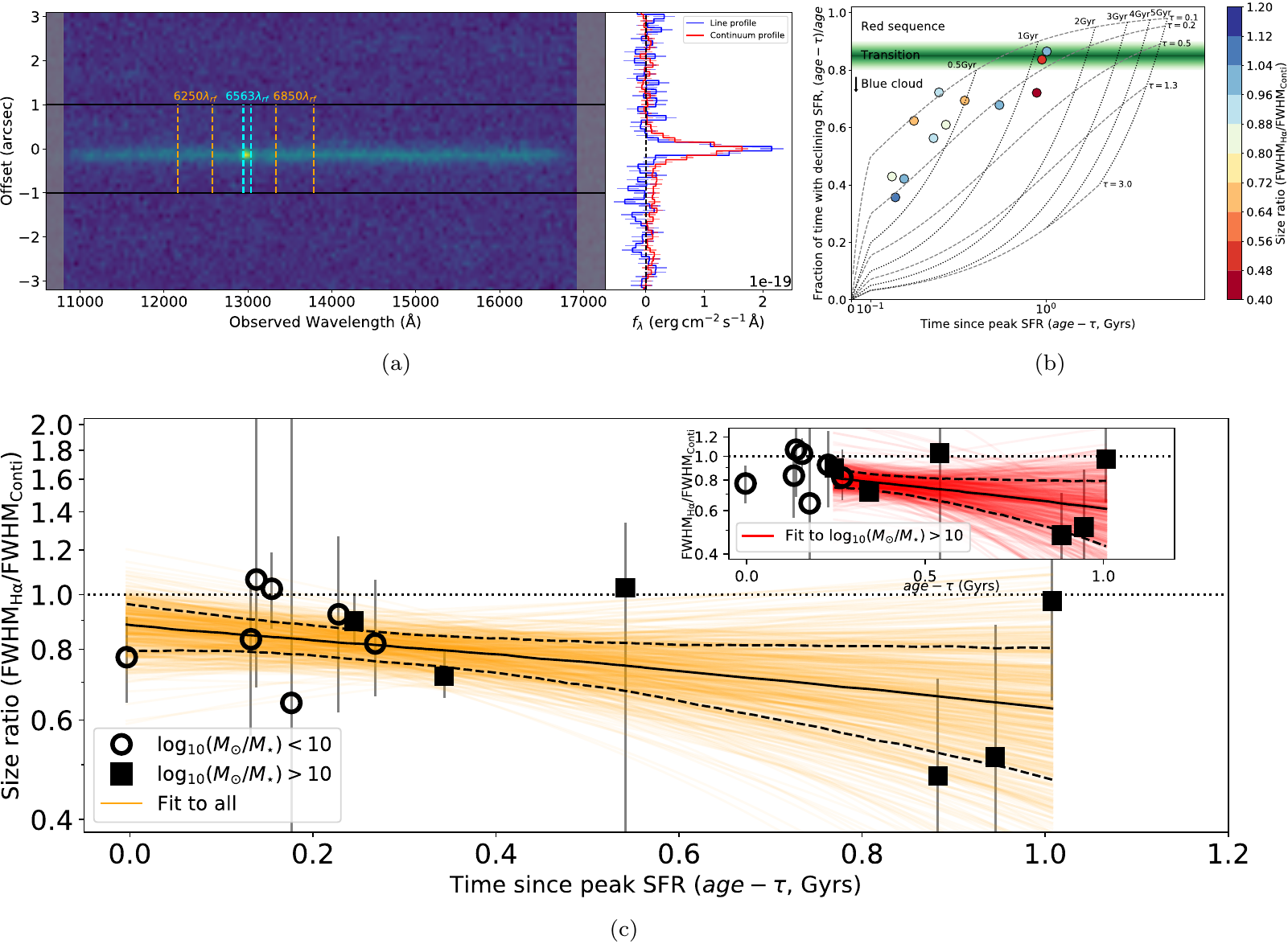}
\caption{(a) 2D G141 grism data and 1D H$\alpha$ (blue) and continuum (red) spatial profiles. (b) Age-$\tau$ diagram, color-coded by the galaxy size-ratios. Galaxies move with time to the right along lines of constant $\tau$ (dashed lines). The green shaded area approximates the transition region between the blue cloud and the red sequence, as determined in \citetalias{Noirot2022}. (c) Size evolution as a function of time since peak SFR.  \label{fig:grismdata}}
\end{figure}

	Figure~\ref{fig:grismdata}(b) shows our sample with respect to the delayed-$\tau$ evolutionary tracks from \citetalias{Noirot2022}, color-coded by the galaxy size-ratios ($\rm FWHM_{H\alpha}/FWHM_{Conti}$). In this diagram, the peak star-formation rate (SFR) of galaxies occurs at $\rm age-\tau=0$, and galaxies move over time from left to right along lines of constant $\tau$ (dashed lines), forming a time sequence as they shut down their star formation 
	and migrate from the blue cloud to the red sequence (see \citetalias{Noirot2022} for details). 
	Two galaxies, at late $\rm age-\tau$ and near the blue-cloud/red-sequence transition region, seem to show smaller H$\alpha$-to-continuum size-ratios than the rest of the sample.
	 
	Figure~\ref{fig:grismdata}(c) shows the evolution of the size ratio with respect to the time since the peak of star-formation ($\rm age-\tau$, in Gyrs) for our fast-$\tau$ population. 
	While the uncertainties on the measurements are relatively large, most galaxies seem to have H$\alpha$-to-continuum size-ratios consistent or close to unity. Only three sources (one with $\rm age-\tau\sim0.4$~Gyrs, and two with $\rm age-\tau\sim0.9$~Gyrs) seem to statistically deviate from size ratios of unity. We fit the size -- ($\rm age-\tau$) relation with a first order polynomial using the {\tt python} Markov Chain Monte Carlo package {\tt emcee} (\citealp{Foreman2013}) for all galaxies (orange curves in the Figure), and for galaxies with $\log_{10}(M_{\star}/M_{\odot}) > 10$ only (red curves, inset). As shown, we find a trend consistent with some sources possessing H$\alpha$ sizes smaller than the stellar continua as galaxies age and progressively shut down their star formation.

\section{Discussion \& Conclusions} \label{sec:conclusion}

	The H$\alpha$-to-continuum size-ratios of the $z=1.0-1.4$ rapidly-quenching galaxies in our sample are mostly consistent with unity (similar line and continuum sizes), except for a few sources with small H$\alpha$ sizes compared to the stellar continua. One of these sources is at a similar evolutionary stage as the bulk of our sample. The two sources with the smallest size ratios in our sample are however seen at a later stage of their evolution. Overall, this suggests that while the past and present star-formation is still co-located for rapidly-quenching galaxies closer to their peak of star-formation, as they age further and move towards quiescence, the remaining star-formation in some (but potentially not all) of these galaxies becomes more centrally concentrated compared to their older stellar components.
	
	While our sample is too small and uncertainties too large to be definitive on whether outside-in quenching mechanisms are the main driver of rapid galaxy evolution/quenching, we note that the trend seen in this work is intriguing and that the approach shows potential. We plan to revisit this topic with the {\it JWST} GTO program CANUCS (the Canadian NIRISS Unbiased Cluster Survey; \citealp{Willott2022}). 
	
	Note that the advantage of the methodology presented here is to extract both the line and the continuum spatial profiles from the same grism spectroscopic data and to evaluate the continuum profiles within small wavelength windows around the emission lines. This offers an alternative to approaches measuring the line and continuum profiles from disjoint datasets (i.e., from the grism spectroscopy and the broad-band imaging, respectively), and also removes the need for modelling and subtracting the grism continuum model from the grism data to isolate the line contribution. This approach does not suffer from using the associated broad-band imaging to measure the stellar continuum size which might also contain, and therefore be correlated to, the emission line contribution.
	However, note that the current approach might present biases from measuring spatial profiles on combined, drizzled grism data which include visits with different dispersion directions. This potential bias would need to be evaluated further before generalizing this method to larger datasets.

%
%

%



\software{
{\tt ASTROPY} \citep{Astropy2018}, {\tt MATPLOTLIB} \citep{Hunter2007}, {\tt NUMPY} \citep{Harris2020}, {\tt SCIPY} \citep{Virtanen2020}.
          }



%
%


\bibliography{references}{}
\bibliographystyle{aasjournal}



\end{document}